\documentclass[sort&compress
    ,final            
    ,numbers              
  ]
  {aipproc}

\usepackage{amssymb,amsmath,amstext,amsthm,amsfonts}
\layoutstyle{6x9}

\usepackage[usenames]{color}
\definecolor{magenta}{cmyk}{0,1,0,0}

\newcommand{\plotscale}{.65}

\begin{document}

\title{Nuclear Effects in Generators: the Path Forward}


\author{Ulrich Mosel}{
  address={Institut fuer Theoretische Physik, Universitaet Giessen, D-35392 Giessen, Germany}
}

\begin{abstract}
The extraction of neutrino oscillation parameters requires the determination of the neutrino energy from observations of the hadronic final state. The use of nuclear targets then requires the use of event generators to isolate the interesting elementary processes and to take experimental acceptances into account. In this talk I briefly summarize the history of event generators and their use in nuclear physics, talk briefly about the generators used in the neutrino community and then discuss future necessary developments.

\end{abstract}

\maketitle


\section{Introduction}
The extraction of neutrino oscillation parameters from long-baseline experiments requires the knowledge of the neutrino energy. Since -- due to their production mechanism -- the neutrino beams are not monoenergetic, but instead contain a wide range of energies, the neutrino energy has to be reconstructed from the final state of the reaction. This can be done by identifying a certain, fixed reaction type and using quasifree kinematics, assuming that the reaction happens on a nucleon at rest. This is the method used by experiments such as MiniBooNE and K2K. Another possibility is to determine the  full energy of the final state by calorimetric measurements; this method is used by both the MINOS and OPERA detectors. In case of nuclear targets the first method suffers from the fact that the actual reaction of a neutrino with a nucleon is not directly observable; final state particles, nucleons and mesons, reach the detector only after sizeable final state interactions (fsi). This poses a problem in, first, identifying a particular initial reaction (usually quasielastic (QE) scattering) and, second, in determining the final state energies of the first interaction from the asymptotic particles that have experienced fsi. In addition a natural, always-present smearing due to Fermi-motion must be taken into account \cite{MoselNUINT}. In the calorimetric method one has to deal with detection thresholds and, in general, detection efficiencies for all possible hadrons; the latter is also the case for the QE-identifying experiments that also are sensitive to detector acceptances. These known deficiencies in the data have to be corrected for by means of event generators \cite{MoselNUINT,Harris:2004iq} that introduce a sizeable dependence of the final observables on the quality of the special generator used \cite{FernandezMartinez:2010dm}. A detailed comparison of predictions of generators both for QE scattering and for pion production performed for NUINT09 has revealed major differences between the various generators \cite{Boyd:2009zz}.

\section{Event Generators in Nuclear Physics}
Since the main complications in identifying elementary neutrino processes in nuclear targets come from nuclear final state interactions and since these are independent of the specific nature of the first, primary interaction it is worthwhile to look for related generator developments for use in quite different reactions in Nuclear Physics.
Precursors of present-day
transport theoretical implementations were Monte-Carlo (MC) generators written to
describe the final-state interactions in reactions of nucleons with
nuclear targets. While the earliest suggestion of such methods goes
back to Serber~\cite{Serber:1947zza}, the method has then been picked
 up by, among others, Metropolis et al{}~\cite{Metropolis:1958sb},
Bertini~\cite{Bertini:1963zz}, and Cugnon~\cite{Cugnon:1980zz}. About 20 years ago, with the advent of heavy-ion reactions, first methods were
developed to describe the dynamical evolution of a colliding
nucleus-nucleus system
\cite{Bertsch:1984gb,Stoecker:1986ci,Bauer:1986zz,Bertsch:1988ik,Cassing:1990dr,Danielewicz:1991dh}
while taking into account the hadronic potentials and the equation of
state of nuclear matter within the Boltzmann-Uehling-Uhlenbeck (BUU)
theory.  These codes thus went beyond the simple MC generators used
until then that could not account for effects of potentials on the
propagation of particles. In addition, mesons were no longer instantaneously produced and, correspondingly, resonance degrees of freedom with their physical lifetimes were incorporated. The GiBUU code has its origins in this era; it started with
applications to heavy-ion reactions (see
\cite{Bauer:1986zz,Cassing:1990dr,Teis:1996kx} and further references
therein). With the availability of ultra-relativistic heavy-ion beams
at the AGS, the CERN SPS, RHIC, and the LHC this field has flourished
ever since; for a more extended listing of transport codes presently used in heavy-ion
physics see Sect.\ 1 in \cite{Buss:2011mx}.

The first attempt to use transport-theoretical
methods for the description of elementary processes on nuclei was an
investigation of inclusive pion-nuclear reactions
\cite{Salcedo:1987md} in an MC calculation. A first calculation within
the framework of a BUU theory of pion and $\eta$ photo-production off
nuclei was performed by the Giessen group in \cite{Hombach:1994gb,Effenberger:1996rc}
with the same code that had been originally developed to describe
heavy-ion collisions. Since then BUU theory has been used by that
group to analyze a wide class of nuclear reactions involving -- besides heavy ions --
elementary projectiles such as hadrons or electrons, photons, and
neutrinos \cite{Buss:2011mx}. For studies of in-medium properties of hadrons \cite{Leupold:2009kz} it turned out to be
essential to incorporate the transport of off-shell particles.

All the codes and methods mentioned were developed from the outset to describe nuclear phenomena.
Not any special interactions were at the center of interest, but instead the treatment of nuclear many-body reactions and phenomena in a wide range of different reactions. Very independently of these developments specialized event generators for neutrino physics have evolved. Widely used codes such as NUANCE, NEUT, GENIE or new developments such as NUWRO (for a short review of neutrino generators see \cite{Gallagher:2009zza}) have been developed with very little coupling to the nuclear physics or the transport-theoretical communities. This is quite astonishing since the nuclear effects present the greatest challenge to the interpretation of neutrino-nucleus experiments.  Only recently detailed studies of nuclear final state interactions, so far concentrated on $\pi-A$ reactions, have begun \cite{Dytman:2009zza,Dytman:2009zzb}.

\section{Future of generators}
In the following subsection I take up a few points of present interest that have to be improved in all available generators for neutrino-induced reactions on nuclei.

\paragraph{Many-particle interactions}
The observation by MiniBooNE that the cross section identified as being that of quasielastic scattering is significantly larger than the QE cross section obtained by various different theories and generators when they all use the same, world-average axial mass $M_A \approx 1 $GeV (see Fig.\ 15 in \cite{AguilarArevalo:2010zc}) has triggered many discussions and even a call for a new paradigm \cite{Benhar:2010nx}.  Possibly connected with this surplus in QE scattering is the very large pion production cross section obtained in that experiment \cite{Leitner:2009de}.

Since processes connected with initial interactions with two nucleons play a role in reactions with electrons on nuclei \cite{Gil:1997bm} it is natural to expect their influence also in neutrino-induced reactions. Indeed, there is now the widespread belief that the $\approx 25\%$ surplus of events labeled as QE in the MiniBooNE experiment over theoretical calculations for true QE scattering is due to such so-called $2p-2h$ events \cite{Martini:2009uj,Nieves:2011pp,Nieves:2011yp}. Thus, the widely discussed result of a significantly higher axial mass of $M_A \approx 1.2 - 1.3$ GeV extracted from the experiments \cite{AguilarArevalo:2010zc,Gran:2006jn,AguilarArevalo:2010cx} or even $M_A = 1.6$ GeV in \cite{Benhar:2010nx} could then simply be a consequence of trying to force a description of $2p-2h$  contributions by a $1p-1h$ model. So far none of the event generators used by the experiments to convert QE-like into QE events takes these $2p-2h$ processes into account. It is also a theoretical challenge to continue the theoretical treatment of $2p-2h$ contributions to higher energies with an aim to understand why they do not seem to contribute in the energy range of the NOMAD experiment \cite{Lyubushkin:2008pe}.

\paragraph{Knock-out nucleon numbers and multi-particle interactions}
Data on knock-out nucleons are directly measurable and do not require the use of a generator for the definition of the reaction process.
Naively one expects the number of knocked-out nucleons to be an indication for the relative importance of QE scattering vs.\ other processes, since QE scattering (in vacuo) is connected with only one outgoing nucleon. However, due to fsi in a nuclear target  more than one nucleon can be present in the final state even if the initial interaction was with one nucleon only.
\begin{figure*}[tbp]
  \includegraphics[scale=\plotscale]{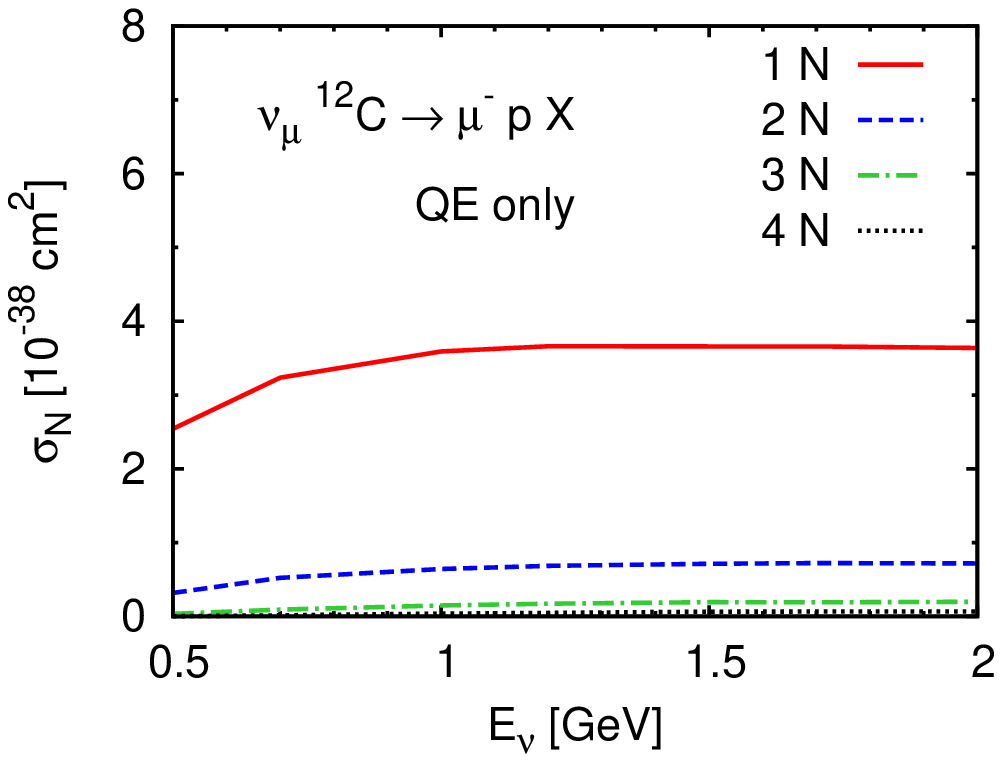}
  \includegraphics[scale=\plotscale]{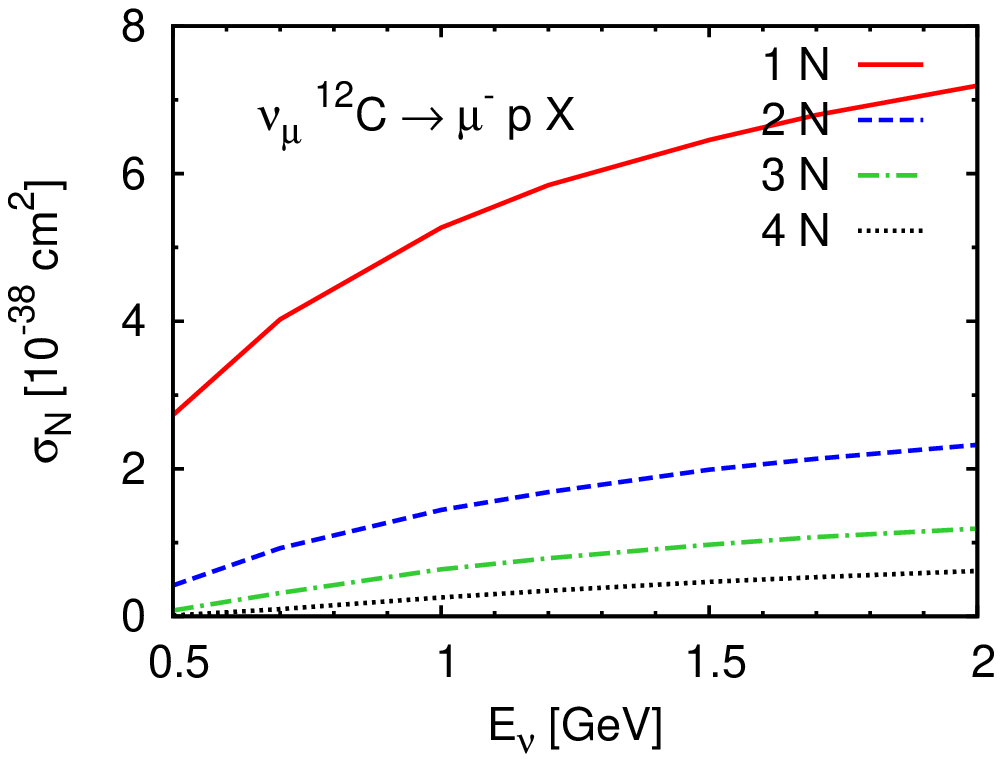}
  \caption{(Color online) Cross section for knock-out of various numbers of nucleons from a $^{12}C$ target as a function of neutrino energy, calculated with GiBUU. The picture on the left shows the result obtained from primary QE events only, the right picture contains the result obtained from all primary interactions (from \cite{Leitnerprivcomm}). \label{fig:numberofN}}
\end{figure*}
Fig.\ \ref{fig:numberofN}, taken from \cite{Leitnerprivcomm} and based on calculations with GiBUU, shows this quite well. The figure on the left shows the cross sections for multiparticle knock-out if the initial interaction is only QE scattering (which is not directly observable on a nuclear target). The figure on the right shows the actually observable cross section for $1 \ldots 4$ nucleons. The cross section for seeing more than one nucleon in the final state increases as a function of neutrino energy and the probability to see 2 nucleons amounts to 20\% of the 1-nucleon contribution at 1 GeV even though the initial, first interaction always happened on one nucleon only.

It is to be expected that this number will increase if the $2p-2h$ contributions that can amount to about 25\% in the total cross section are taken into account in the generators. It is clearly necessary to implement such processes explicitly in any event generator since the many particle final states will affect the energy reconstruction. The recent proposal by Bodek et al.\ \cite{Bodek:2011ps} to include such processes simply by increasing the single-particle transverse response misses this important physics since it leads to incorrect knock-out nucleon numbers in the final state of the initial interaction.

The latter is also true when treating effects of groundstate correlations \cite{Benhar:1989aw,CiofidegliAtti:1990vn} in the impulse approximation (IA) where the first interaction happens on one (dressed) nucleon only. The spectral functions used there contain the effects of these correlations in terms of nontrivial connections between energy and momentum of the nucleons, i.e.\ spectral functions. There are, however, many-body contributions, for example those connected with exchange currents, that cannot be taken care off by an IA using single particle spectral functions. These have to be added explicitly into event generators.

 For handling these off-shell effects in the spectral functions in the very first, primary interaction there are methods (recipes) available that have been developed over the last 30 years \cite{Bodek:1981wr,Benhar:2006wy}; these methods seem to be accepted in the community, but suffer from problems connected with insufficient knowledge of off-shell properties of the interaction vertices and gauge invariance. By assumption they furthermore  do not allow for production and further propagation of off-shell nucleons. This restriction is hard to justify in the presence of fsi. Based on the seminal work of Kadanoff and Baym \cite{KadanoffBaym} developed around 40 years ago practical implementations of transport of off-shell particles have become available only during the last decade. While GiBUU does contain this new physics, all other transport codes used in neutrino physics do not. That off-shell transport can make a difference for knock-out nucleon cross sections and particle production close to threshold was shown in \cite{LeitnerDiss,Bertsch:1995ig}.

\paragraph{Broad-Band Generators}
Various reaction mechanisms, such as true QE, pion production, resonance excitation, DIS and $2p-2h$ contributions dominate in different regimes of the neutrino energy so that any broad-band experiment necessarily contains contributions from different reaction mechanisms.
 To isolate quasielastic scattering, for example, for use in the neutrino energy reconstruction then requires the use of event generators and theories which are reliable not only for QE scattering, but are well tested and reliable for a broad class of relevant reactions. In particular, the errors connected with the use of these generators have to be well under control \cite{Harris:2004iq,FernandezMartinez:2010dm}. The latter can be assessed only by comparison with results from different experiments.

\paragraph{Verification of Generators}
There are lots of data about photo- and electro-production of mesons on nuclear targets, in the energy range of a few 100 MeV up to 200 GeV  which cover the energy regime relevant for present day's neutrino experiments. These reactions involve a closely related primary interaction and the very same fsi as in the neutrino-induced reactions. They could thus be used to check the reliability and accuracy of the generators used. While it seems that some of such checks were done (but not published) with NEUT \cite{Hayato:2002sd} none have been done with the other standard neutrino generators; on the contrary, GiBUU was developed to handle such interactions \cite{Buss:2011mx}. The necessary code changes are minimal (only the initial axial coupling has to be turned off). Such comparisons are ultimately more meaningful than just comparisons of generators with other generators.

\section{Summary}
Mainly because of the presence of high-statistics data obtained during the last few years at K2K, MiniBooNE (and soon T2K, Minerva and NOvA) our understanding of reactions of neutrinos with nuclei has improved a lot. It seems to be quite clear now that the initial analyses that led to very different, higher values of the axial mass than the accepted world-average value of around 1 GeV were incorrect because they tried to force a description of $2p-2h$ effects in the primary interaction in terms of a one-body model. Obvious is now that all generators have to be extended to include the $2p-2h$ effects; what is open is still the question if these effects are unambiguously observable and how their inclusion will affect the energy reconstruction.
 Since these two-body couplings are well known from interactions of real or virtual photons with nuclei the lesson to be learned from this recent development is that neutrino generators do not only have to be compared with each other. Foremost, they have to be compared with electron- or photon-induced data on nuclear targets since here the primary interaction is quite similar to that for neutrinos and the final state interactions are the same. Very few, if none, of such comparisons exist for the 'working-horse generators'.

 Finally, it is clear that the extraction of neutrino oscillation parameters depends to a large extent on the reliability and quality of the event generators used. The quality of these generators could only benefit from a much closer contact to nuclear theory and to other generator developments in nuclear physics. This is still a long way to go for neutrino generators until they reach the level of sophistication that the experimental equipment in these long baseline experiments already has.


\begin{theacknowledgments}
I am grateful to Tina Leitner, Olga Lalakulich and Luis Alvarez-Ruso for many helpful discussions on neutrino-nucleus interactions and their theoretical description.

This work was supported by Deutsche Forschungsgemeinschaft and HIC for FAIR.
\end{theacknowledgments}

\end{document}